\documentstyle[12pt]{article}

\begin{document}

\pagestyle{empty}

\begin{flushright}
{\bf McGILL-00-02}\\
{\bf UA/NPPS-01-00}
\end{flushright}

\vglue 0.3 cm

\begin{center} \begin{Large} \begin{bf} 
Polarized Photoproduction of Heavy Quarks in Next-to-Leading Order$^{+}$
\end{bf} \end{Large} \end{center}
\vglue 0.35cm
{\begin{center} 
A.P.\ Contogouris$^{a,b,1}$,
Z.\ Merebashvili$^{a,*,2}$ and 
G.\ Grispos$^{b,3}$ \end{center}}
\parbox{6.4in}{\leftskip=1.0pc
{\it a.\ Department of Physics, McGill University, Montreal,
Qc., H3A 2T8, Canada}\\
{\it b.\ Nuclear and Particle Physics, University of Athens, 
Athens 15771, Greece}\\
\vglue -0.25cm
}
\begin{center}
\vglue 1.0cm
\begin{bf} ABSTRACT \end{bf}
\end{center}
{

The results of a next-to-leading order calculation of heavy quark production 
in longitudinally polarized photon-nucleon collisions are presented.
At c.m. energy $\sqrt{S}=10$ GeV, for $\vec \gamma +\vec p \rightarrow c+X$,
cross sections differential in the transverse momentum and rapidity of the
charmed quark $c$ and the corresponding asymmetries are presented;
also, as functions of $\sqrt{S}$, integrated cross sections, $K$-factors
and the corresponding asymmetries are given. Errors in the
asymmetries are estimated and the possibility to distinguish between
three scerarios differing essentially in the polarized gluon distribution
is discussed.
}
\renewcommand{\thefootnote}{+}
\footnotetext{Also supported by the Secretariat of Research and
Technology of Greece and by the Natural Sciences and Engineering Research
Counsil of Canada.}
\renewcommand{\thefootnote}{*}
\footnotetext{Present address: High Energy Physics Institute,
Tbilisi State University, University St. 9, 380086 Tbilisi, 
Republic of Georgia.}
\renewcommand{\thefootnote}{\arabic{footnote}}
\addtocounter{footnote}{1}
\footnotetext{e-mail: apcont@physics.mcgill.ca, acontog@cc.uoa.gr}
\addtocounter{footnote}{1}
\footnotetext{e-mail: mereb@sun20.hepi.edu.ge}
\addtocounter{footnote}{1}
\footnotetext{e-mail: ggrispos@cc.uoa.gr}

\newpage

\pagestyle{plain}
\setcounter{page}{1}

In spite of the welth of data on polarized deep inelastic scattering,
the size and shape of the polarized gluon
distribution $\Delta g(x) $ remains a central problem in Spin Physics.
Important progress needs experiments on processes with longitudinally
polarized initial particles dominated by subprocesses with initial gluons.
Such a process is
\begin{equation}
\vec{\gamma}+\vec{p}\rightarrow Q \bar{(Q)} + X,
\end{equation}
where $Q \bar{(Q)}$ denotes heavy quark (antiquark), and, in general,
it is
dominated by the subprocess
\begin{equation}
\vec{\gamma}+\vec{g}\rightarrow Q + \bar{Q}
\end{equation}
Proposals on experiments closely related to, or in, (1) exist in various stages
of approval $[$\ref{r1}$]$.

On the other hand, the importance of determining higher order corrections (HOC)
is well known. In this Letter we report the essential results of a
calculation of the (next-to-leading) HOC.

The subprocesses contributing to (1) are as follows:
\begin{itemize}
\item[(A)]
\underline{At leading order} (LO, $\alpha \alpha_s$) $[$\ref{r2}, \ref{r3}$]$
\item[(A1)]
The Born subprocess of (2).
\item[(A2)]
The resolved $\gamma$ via $\vec{q}\vec{\bar{q}}\rightarrow Q \bar{Q}$
and $\vec{g}\vec{g}\rightarrow Q \bar{Q}$. These involve the polarized photon
structure functions $\Delta F_{q/ \gamma}$ and $\Delta F_{g/ \gamma}$,
known only theoretically; hopefully, more information will eventually
come from experiments on (1).
\item[(B)]
\underline{At next-to-leading order} (NLO, $\alpha \alpha_s ^2$)
\item[(B1)]
The loop and Bremsstrahlung (Brems, i.e.
$\vec{\gamma}\vec{g}\rightarrow Q \bar{Q}g$) associated with (2).
\item[(B2)]
The subprocess
$\vec{\gamma}\vec{q}\rightarrow Q \bar{Q}q$
\end{itemize}
At NLO, scheme independent cross sections can be rigorously obtained
only for the sum of resolved and direct contributions.

Notice, that the Abelian part of (B1) provides the HOC to
\begin{equation}
\vec{\gamma}\vec{\gamma}\rightarrow Q \bar{Q};
\end{equation}
these HOC have been determined $[$\ref{r4}, \ref{r5}$]$. HOC to the process (3)
are of interest in themselves: In searches of the Higgs boson (mass $m_H$),
for $90$ GeV$\leq m_H \leq 2m_W$, in future $\gamma -\gamma$ colliders,
the dominant decay mode is $H \rightarrow b \bar{b}$. With polarized
$\gamma$' s, the Born contribution to the background
$\vec{\gamma}\vec{\gamma}\rightarrow b \bar{b}$
is much suppressed; however, due to gluon Brems, HOC have an important
effect $[$\ref{r4}, \ref{r5}$]$.
 
A calculation of NLO corrections for (1) already exists $[$\ref{r6}$]$.
Our work, however, makes use of a different regularization scheme; also,
in the treatment of the
soft and collinear contributions, contrary to $[$\ref{r6}$]$ which separates 
them from the hard parts
via a cut parameter, we apply more conventional methods $[$\ref{r4}$]$.
Thus, in view of the importance of (1), we believe that an independent
calculation is worthwhile. Comparisons, as much
as possible, with the results of $[$\ref{r6}$]$ will be also reported.

Most conveniently, singularities are eliminated by dimensional methods.
For polarized reactions this requires extension of the Dirac matrix
$\gamma_5$ in $n=4-2\varepsilon$ dimensions. There are several schemes for
this, and, as in $[$\ref{r4}$]$, we follow that of dimensional reduction.
This scheme violates the Ward identity between the vertex and quark self
energy functions, but care has been taken by introducing a (finite)
counterterm, as discussed in $[$\ref{r4}$]$.
The wave function and mass renormalizations are carried on shell $[$\ref{r4}$]$.

In the present case charge renormalization is also required. We introduce
\begin{equation}
A_{\varepsilon } \left( m\right) =
\left( \frac{g}{4\pi }\right)^2 \left( \frac{4\pi \mu ^2}{m^2 }\right)
^{\varepsilon }  \Gamma \left( 1+\varepsilon \right) 
\end{equation}
where $\mu$ is an arbitrary mass scale entering in $n=4-2\varepsilon$
dimensions via the change $g \rightarrow g\mu ^{\varepsilon }$. Let
$N_{lf}=$ number of light flavors and $b\equiv \left( 11N_c
-2N_{lf}\right) /6$. Then we carry charge renormalization by introducing
the counterterm
\renewcommand{\theequation}{\arabic{equation}a}
\setcounter{equation}{3}
\begin{equation}
-\frac{1}{\varepsilon } \left( A_{\varepsilon } \left( \mu_R\right) b
-\frac{1}{3} A_{\varepsilon } \left( m\right) \right)
\end{equation}
where $\mu_R$ is a regularization mass. In this scheme graphs containing
internal loops of the heavy quark $Q$ are subtracted out, so that $Q$ is
decoupled. This is consistent with parton distributions of which the
evolution is determined from split functions involving only light quarks,
as is our case.

Finally, the $gQ \bar{Q}$ vertex was renormalized via the Slavnov-Taylor
identities $[$\ref{r7}$]$.
\renewcommand{\theequation}{\arabic{equation}}
\setcounter{equation}{4}

\vglue 0.5 cm

Now, loop contributions to (B1) were determined via Passarino-Veltman
techniques $[$\ref{r8}$]$;
and $2\rightarrow 3$ parton contributions by going to the c.m.
(Gottfried-Jackson) frame of $\bar{Q} (Q)$ and final $g$ $[$\ref{r9},
\ref{r4}$]$. Also, (B2) was treated by going to the c.m. frame of
$\bar{Q} (Q)$ and final $q$. The integrals listed in the
Appendices A and C of $[$\ref{r9}$]$
have been very useful. Some remaining integrals are given in an Appendix
of $[$\ref{r10}$]$.

Singularities $\sim 1/\varepsilon ^2$ appearing in the cross sections of
(B1) are cancelled by adding
loops and Brems. The cancellation of singularities $\sim 1/\varepsilon $
appearing in (B1) and (B2)
required the addition of factorization counterterms. With $p_1, p_2, p_3$
the 4-momenta of $\gamma$,
initial parton and observed $Q (\bar{Q})$, define 
\[
s=(p_1+p_2)^2, \qquad  t=(p_3-p_1)^2-m^2, \qquad u=(p_3-p_2)^2-m^2. 
\]
Our counterterms are defined in the $\overline{MS}$ scheme and have the
general form:
\begin{equation}
\Delta \frac{d\sigma_{cter}}{dtdu}=\frac{1}{\varepsilon }
\frac{K(\varepsilon )}{s} \Delta P_{ab}(x)
\left( \frac{m^2}{M^2}\right)^{\varepsilon } \Delta \frac{d \hat{\sigma}_B}{dt} 
\end{equation}
where $\Delta P_{ab}(x)$ split function (in $n=4$ dimensions, see below),
$x$ proper dimensionless variable,
$M$ the factorization scale, $K(\varepsilon )$ a kinematic factor determined
from phase-space and color and $\Delta d\hat{\sigma}_B /dt$ the Born
cross section of a $2\rightarrow 2$ subprocess with $s$ and either $t$ or
$u$ replaced by $xs$ and either $xt$ or $xu$. For
$\vec{\gamma}\vec{g}\rightarrow Q \bar{Q}g$ the
$2\rightarrow 2$ subprocess is (2);
for $\vec{\gamma}\vec{q}\rightarrow Q \bar{Q}q$
one needs two counterterms, one involving (2) and another involving
$\vec{\bar{q}}\vec{q}\rightarrow Q \bar{Q}$ $[$\ref{r11}$]$. 

The resulting finite cross sections are convoluted with polarized parton
distributions whose evolution is determined by 2-loop anomalous dimensions
$[$\ref{r12}, \ref{r13}$]$.
On their basis, several groups have constructed sets of such
distributions, differing mainly in the shape and size of $\Delta g$. We
use throughout the NLO sets of one group $[$\ref{r14}$]$. Also, we use the
NLO expression of the running coupling constant
$\alpha _s (\mu)$ with $\Lambda =231$ MeV and $N=N_{lf}+1=4$ flavours.

The scheme for extending $\gamma _5$ in $n$ dimensions used in
$[$\ref{r12}, \ref{r13}$]$ is not dimensional reduction, so the addition
of certain conversion terms is necessary.
The form of the conversion terms is easily given in terms of Eq. (5): In
$n$ dimensions the split functions have the form:
\begin{equation}
\Delta P_{ba} ^n \left( x, \varepsilon \right)=\Delta P_{ba} \left( x \right)+
\varepsilon \Delta P_{ba} ^{\varepsilon } \left( x \right);
\end{equation}  
it is $\Delta P_{ba}^{\varepsilon } \left( x \right)$ that depends on the
scheme. The conversion terms are determined from the difference of $\Delta
P_{ba} ^{\varepsilon } \left( x \right)$ in the
different schemes. In dimensional reduction
$\Delta P_{ba} ^{\varepsilon } \left( x \right) \left(= P_{ba}
^{\varepsilon } \left( x \right) \right)$ $= 0$.
Refs.~$[$\ref{r12}, \ref{r13}$]$ use the t' Hooft-Veltman scheme,
modified so that
$\Delta P_{qq} ^n \left( x, \varepsilon \right)=P_{qq} ^n \left( x, \varepsilon \right)$;
then $\Delta P_{ab} ^{\varepsilon } \left( x, \right) \neq 0$. In terms of these
$\Delta P_{ab} ^{\varepsilon } \left( x \right)$, the conversion terms are
\begin{equation}
\frac{d\sigma_{conv}}{dtdu}=-
\frac{K(0)}{s} \Delta P_{ab} ^{\varepsilon }(x)
\frac{d \hat{\sigma}_B}{dt} 
\end{equation}  
where $d\hat{\sigma}_B /dt$ the Born cross section of (5) scaled in the same way.

For the polarized parton distributions we use the sets A, B and C of Ref. 14 and for
the unpolarized the most recent version CTEQ5 $[$\ref{r15}$]$. 

Finally, in the absence of any experimental information, as an estimate of the
resolved $\gamma$ contributions, we have used the maximal and minimal
saturation LO sets of $\Delta F_{q/ \gamma}$ and  $\Delta F_{g/ \gamma}$
of $[$\ref{r16}$]$,
as well as the set of the asymptotic solutions $[$\ref{r17}$]$. In brief,
those of $[$\ref{r17}$]$
give the largest contributions, whereas those of the minimal saturation
set give the smallest.

The analytical calculations were carried with REDUCE and to some extent
with FORM.

\vglue 0.5 cm

Subsequently we present results for $Q=$c-quark with $m=1.5$ GeV only at
$\sqrt{S_{\gamma p}}=\sqrt{S}=10$ GeV, relevant to the experiments (a)
and (b) of 
$[$\ref{r1}$]$. Higher energies and $Q=$b-quark are considered elsewhere
$[$\ref{r10}$]$.
Also, in relation with (4a), we take $\mu_R=\mu$.   

Fig.~1 presents results related with the differential cross sections
$\Delta d\sigma /d p_T$
versus $x_T\equiv 2p_T/ \sqrt{S}$, where $p_T$ the transverse momentum of $Q$;
measurement of such cross sections is possible in (b) of $[$\ref{r1}$]$.
Subsequently we denote by $\Delta d\sigma_B /d p_T$, $\Delta
d\sigma_{res} /d p_T$,
and $\Delta d\sigma_{\gamma q}  /d p_T$ the contributions
to the physical cross section of (A1), (A2), and (B2) correspondingly,
and by $\Delta d\sigma /d p_T$
that of the sum (A1), (B1) and (B2). We use the
scale $\mu =M= \left( p_T^2 +m^2 \right) ^{1/2}$; the stability of our
results against
variations of $\mu$ and $M$ is studied in $[$\ref{r10}$]$.
In Fig.~1(a) the cross sections $\Delta d\sigma /d p_T$ and the
corresponding $\Delta d\sigma_B /d p_T$ (denoted by a $*$) are determined
for sets A, B and C of $[$\ref{r14}$]$.
The presented $\Delta d\sigma_{res} /d p_T$ and $\Delta d\sigma_{\gamma
q}  /d p_T$ correspond to set B and the former to the maximal
saturation set of $[$\ref{r16}$]$.
In calculating $\Delta d\sigma_B /d p_T$ and $\Delta d\sigma_{res} /d p_T$ 
we use the NLO sets of $[$\ref{r14}$]$.$^{(a)}$

\renewcommand{\thefootnote}{(a)}
\footnotetext{In this way the effect of the perturbative HOC, as it is
reflected e.g. in the magnitude of $K$-factors (see Eqs.~(10)), is made
more clear. This is particularly true for polarized reactions, in which
the LO and NLO $\Delta g$ differ significantly.}

Fig.~1(b) presents the asymmetries
\begin{equation}
A_{LL} (p_T)= \frac{\Delta d\sigma /d p_T}{d\sigma /d p_T}
\end{equation}
for sets A, B and C. The resolved $\gamma$ contributions have been left
out in view of their smallness and of the fact that the stage of their
present knowledge does not permit a scheme independent calculation.
Following usual practice in calculating $d\sigma /d p_T$
we average over $n-2$ spin degrees of freedom for every incoming boson.
Finally, the errors in Fig.~1(b) have been estimated from:
\begin{equation}
\delta A_{LL} =  \frac{1}{P_B P_T \sqrt{L \sigma \epsilon } }
\end{equation}
In (9) we use unpolarized cross section $\sigma$ integrated over a bin of
$x_T$ corresponding to $\Delta p_T=0.5$ GeV and the conditions of
$[$\ref{r1}$a]$ ($P_B=80 \%  $,
$P_T=25 \%  $, $\epsilon  =0.014 $ and $L=2$ $fb^{-1}$).
Note that the proposal $[$\ref{r1}$b]$ amounts to better conditions and
thus smaller $\delta A_{LL}$.

On the basis of our calculations and Fig.~1 we remark the following:
\begin{itemize}
\item[(i)]
Defining by $\Delta d\sigma_{\gamma g}/ d p_T$ the contribution of (A1)
and (B1) we may introduce the $K$-factors
\begin{equation}
K=\Delta \frac{d\sigma} {d p_T}/ \Delta \frac{d\sigma_B} {d p_T},
\qquad
K_{\gamma g}=\Delta \frac{d\sigma_{\gamma g}} {d p_T}/ \Delta
\frac{d\sigma_B} {d p_T}
\end{equation}
For sets A and B and $x_T \geq 0.3$, where perturbative QCD is more
trustworthy, the $K$-factors are $>1$ and, in particular $K_{\gamma g}$,
fairly large.
The largeness of $K_{\gamma g}$ is partly due to the fact that
$\sqrt{S}=10$ GeV is fairly low, so $\alpha _s (\mu)$ is fairly large.
\item[(ii)]
$\Delta d\sigma_{\gamma q}/ d p_T$ are found to change little between
sets A, B and C; the reason is that the valence distributions $\Delta u$
and $\Delta d$ vary little. Thus Fig.~1(a)
presents $\Delta d\sigma_{\gamma q}/ d p_T$ for only one set (B). 
\item[(iii)]
In general, $\Delta d\sigma_{\gamma q}/ d p_T$ are significantly smaller than
$\Delta d\sigma_{\gamma g}/ d p_T$ for sets A and B. Note that also in
unpolarized photoproduction we find that $\gamma q\rightarrow Q \bar Q q$
contributes much less than
the NLO $\gamma g\rightarrow Q \bar Q $, in full accord with $[$\ref{r18}$]$.
\item[(iv)]
Most importantly, Fig.~1(b) shows that at $x_T \approx 0.4$ one can
distinguish sets A and C and perhaps also all A, B, C. 
\end{itemize}

Fig.~2 presents results related with the distributions $\Delta d\sigma/dY$
where $Y$ the c.m. rapidity of $Q$ with respect to the photon.
The subsequent notation is equivalent to the above and parts (a) and (b)
are equivalent to these of Fig.~1.
Here we use the scale $\mu =M=2m$. The errors in Fig. 2(b) have been
estimated using Eq. (9) with the conditions of $[$\ref{r1}$a]$ and the
unpolarized cross section integrated over a bin $\Delta Y=1$.

On the basis of Fig.~2 we remark:
\begin{itemize}
\item[(i)]
At $Y>0$, $\Delta d\sigma / dY$ for sets A, B are significantly larger
than for C and than
$\Delta d\sigma_{\gamma q} / dY$ and $\Delta d\sigma_{res} / dY$. (Fig.~2(a))
\item[(ii)]
Taking into account the errors, Fig.~2(b) suggests that $Y\approx 1.25\sim
1.5$ is the
best region to distinguish set C from A or B. The region $Y<0$ leads to
large errors because $d\sigma / dY$, like $\Delta d\sigma / dY$ (Fig.
2(a)), is small.
\item[(iii)]
Figs~2(a) and 2(b) suggest that integrating the cross sections in the
range $1\leq Y\leq 1.5$
offers perhaps the best possibility to distinguish A or B form C.
\end{itemize}

Finally, Fig.~3 presents results related with the integrated cross
sections $\Delta \sigma$
versus the c.m. energy $\sqrt{S_{\gamma p}}=\sqrt{S}$ in a range
including $\sqrt{S}=10$ GeV.
This quantity will be measured
in both experiments $[$\ref{r1}$a]$ and $[$\ref{r1}$b]$. Here the scale
is again $\mu =M=2m$.

To show clearly the NLO effects, we present  $K$-factors (Fig.~3(a)) in
addition to integrated cross
sections (Fig.~3(b)). At $\sqrt{S}=10$ GeV the $K$-factors for all sets
exceed $K=1$. In the range
$7< \sqrt{S} <14$ GeV for sets A and B, $K$ are smooth, but for C, $K$ is
discontinuous due to the vanishing of $\Delta \sigma_{B}$ at
$\sqrt{S}\approx 11$ GeV.

Most important is Fig.~3(c), which presents NLO asymmetries. The error at
$\sqrt{S}= 10$ GeV is estimated
using in (9) again the conditions of $[$\ref{r1}$a]$. Under these
conditions our results show that sets A and
C can be distinguished, but not sets A and B or B and C. Perhaps the
proposed SLAC experiment $[$\ref{r1}$b]$, which will give results at
somewhat lower $\sqrt{S}$ and, as stated, amounts to better
conditions, can distinguish also B and C.  

\vglue 0.5 cm

At $\sqrt{S}\approx 10$ GeV of importance is the precise knowledge of the
charmed quark mass $m$. We understand that recently the uncertainity in
$m$ has been somewhat reduced$^{(b)}$.
Using the set B, we find that varying $m$ in the range $1.35 \leq m \leq
1.65$ changes $\Delta \sigma$ in the range $27.11 \leq \Delta \sigma \leq
13.34$ nb and the asymmetry in $6.26 \% \leq A_{LL} \leq 8.88\% $.

\renewcommand{\thefootnote}{(b)}
\footnotetext{We would like to thank A.~Despande for this information.}

Ref.~$[$\ref{r6}$]$ works in the scheme of Ref.~19 and presents most of
its results using the "standard" set of $[$\ref{r20}$]$ (GRSVst);
for integrated cross sections (Fig.~3), it also presents results for sets
A and C of $[$\ref{r14}$]$.
Also, as stated, they treat their soft and collinear gluon parts by the
phase space slicing method of $[$\ref{r18}$]$, which separates them from
the hard gluon parts via a cut parameter.
First, we have checked that our virtual, soft and collinear contributions are in
complete agreement (Appendix C of the first reference of $[$\ref{r6}$]$).
Second, with respect to numerical comparisons, care is needed in
convoluting the LO cross sections with proper (LO or NLO) parton
distributions.
Using also GRSVst, and comparing with their corrected results, regarding 
differential cross sections we find differences not exceeding $3\%$;
however, regarding asymmetries of integrated cross sections,
although in fair agreement, our differences are somewhat larger. 

In conclusion, the COMPASS experiment $[$\ref{r1}$a]$ is expected to
provide useful information on $\Delta g$. This may be said with more
emphasis for the proposed SLAC experiment
$[$\ref{r1}$b]$.
\vglue 1 cm
\begin{center}
{\bf *} 
\end{center}
\vglue 0.5 cm
We thank I. Bojak for providing us with the results for several quantities
we are comparing and for taking his part in doing comparisons.
Thanks are also due to G. Bunce, D. de Florian, B. Kamal and J. K{\"o}rner for
discussions, to W. Vogelsang for discussions and for providing us the sets
of $[$\ref{r20}$]$, to P. Bosted for several communications, to
A.~Despande for useful information and remarks and to V. Spanos and G.
Veropoulos for participating in part of the calculation.

\newpage
\begin{center}\begin{large}\begin{bf}
REFERENCES
\end{bf}\end{large}\end{center}
\vglue .3cm

   \begin{list}{$[$\arabic{enumi}$]$} 
    {\usecounter{enumi} \setlength{\parsep}{0pt} 
     \setlength{\itemsep}{3pt} \settowidth{\labelwidth}{(99)} 
     \sloppy} 
\item \label{r1} 
(a) G. Baum et al, COMPASS Collaboration: CERN/SPLC 96-14 and 96-30;
(b) P. Bosted, SLAC-PROPOSAL-E156, 1997;
(c) W.-D. Nowak, DESY 96-095;
(d) A. de Roeck and T. Gehrmann, DESY-Proceedings-1998-1.
\item \label{r2} 
M. Gluck and E. Reya, Z. Phys. {\bf C39} (1988) 569;
M. Stratmann and W. Vogelsang, Z. Phys. {\bf C74} (1997) 641;
A. Watson, {\it ibid} {\bf C12} (1982) 123.
\item \label{r3} 
B. Lampe and E. Reya, MPI-PhT/98-23 and DO-TH $98/02~^{(+)}$.
\renewcommand{\thefootnote}{(+)}
\footnotetext{This is also the most up to date review of polarized
particle reactions}
\item \label{r4} 
B. Kamal, Z. Merebashvili and A.P. Contogouris, Phys. Rev. {\bf D51} 
(1995) 4808; {\it ibid} {\bf D55} (1997) 3229 (E).
\item \label{r5} 
G. Jikia and A. Tkabladze, {\it ibid} {\bf D54} (1996) 2030.
\item \label{r6}
I. Bojak and M. Stratmanm:
Nucl. Phys. {\bf B540} (1999) 345 and Erratum (to be published); Phys.
Lett. {\bf B433} (1998) 411.
\item \label{r7} 
T. Muta, "Foundations of Quantum Chromodynamics" 
(World Scientific, 1987)
\item \label{r8} 
G. Passarino and M. Veltman, Nucl. Phys. {\bf B160} (1979) 151.
\item \label{r9} 
W. Beenakker et al, Phys. Rev. {\bf D40} (1989) 54.
\item \label{r10}
Z. Merebashvili, A.P. Contogouris and G. Grispos, in preparation.
See also proceedings of  "Spin 99" International Workshop, Prague,
8-12 September 1999.
\item \label{r11}
A.P. Contogouris, S. Papadopoulos and B. Kamal,
Phys. Lett. {\bf B246} (1990) 523.
\item \label{r12} 
R. Mertig and W. van Neerven, Z. Phys. {\bf C70} (1996) 637.
\item \label{r13} 
W. Vogelsang, Phys. Rev. {\bf D54} (1996) 2023;
Nucl. Phys. {\bf B475} (1996) 47.
\item \label{r14} 
T. Gehrmann and W. Stirling, Phys. Rev. {\bf D53} (1996) 6100.
\item \label{r15}
H. Lai et al, Eur. Phys. J. {\bf C12} (2000) 375.
\item \label{r16}
M. Gluck and W. Vogelsang, Z. Phys. {\bf C55} (1992) 353
and {\bf C57} (1993) 309;
M. Gluck, M. Stratmann and W. Vogelsang, Phys. Lett. {\bf B187} (1994) 373.
\item \label{r17} 
J. Hassan and D. Pilling,
Nucl. Phys. {\bf B187} (1981) 563.
\item \label{r18}
J. Smith and W.L. van Neerven, Nucl. Phys. {\bf B374} (1992) 36.
\item \label{r19}
G. t' Hooft and M. Veltman, ibid {\bf B44} (1972) 189.
\item \label{r20} 
M. Gluck, E. Reya, M. Stratmann and W. Vogelsang, Phys. Rev. {\bf D53}
(1996) 4775.
\end{list}

\newpage
\begin{center}\begin{large}\begin{bf}
FIGURE CAPTIONS
\end{bf}\end{large}\end{center}
\vglue .3cm

\begin{list}{Fig.~\arabic{enumi}.} 
    {\usecounter{enumi} \setlength{\parsep}{0pt} 
     \setlength{\itemsep}{3pt} \settowidth{\labelwidth}{Fig.~9.} 
     \sloppy}  
\item{}
Quantities related with the $p_T$-distributions versus $x_T =2p_T /\sqrt{S}$:
(a) Polarized differential cross sections. The LO cross sections are indicated
by $*$. (b) NLO asymmetries for sets A, B and C.
\item{} 
Quantities related with the rapidity distributions: (a) and (b) as in Fig. 1.
\item{} 
Quantities related with the integrated cross sections for sets A, B and C:
(a) Factors $K= \Delta \sigma / \Delta \sigma _{B} $
(b) LO (indicated by $*$) and NLO cross sections. The presented cross section
for $\vec \gamma \vec q \rightarrow Q \bar Q q$ corresponds to set B.
(c) NLO asymmetries.
\end{list}
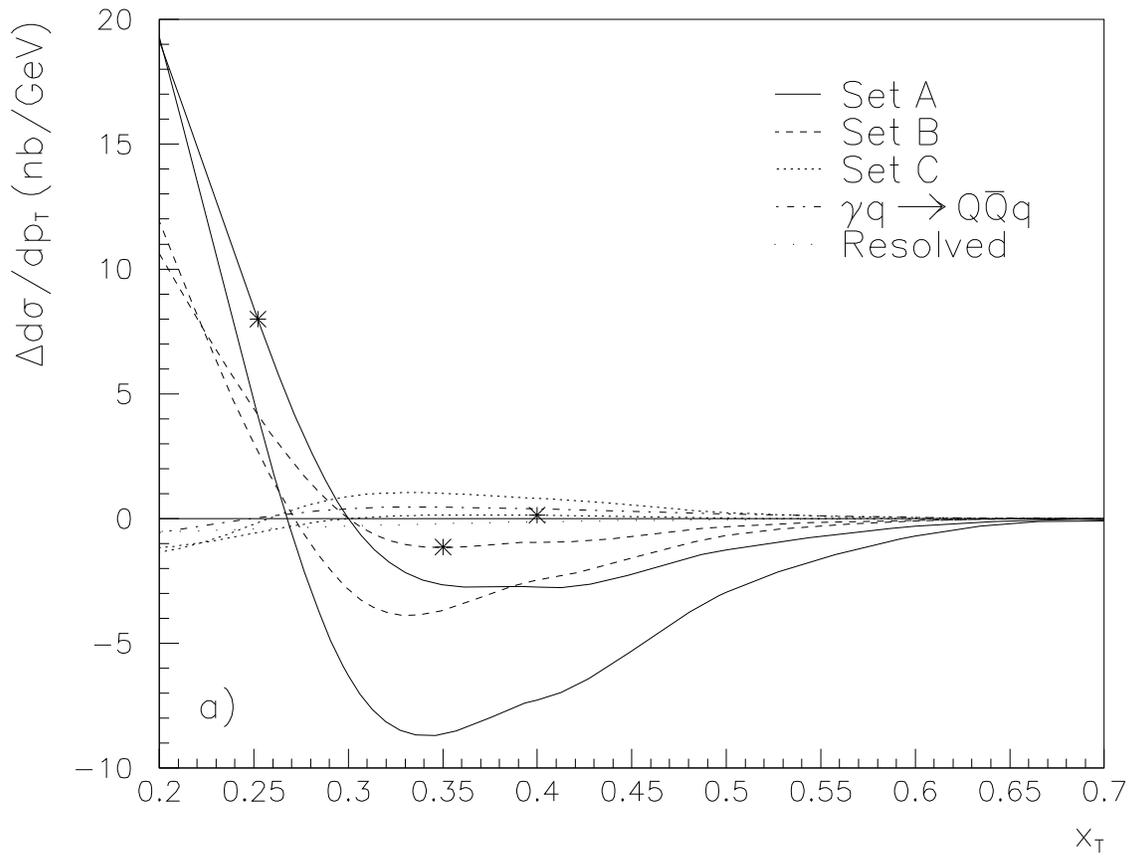
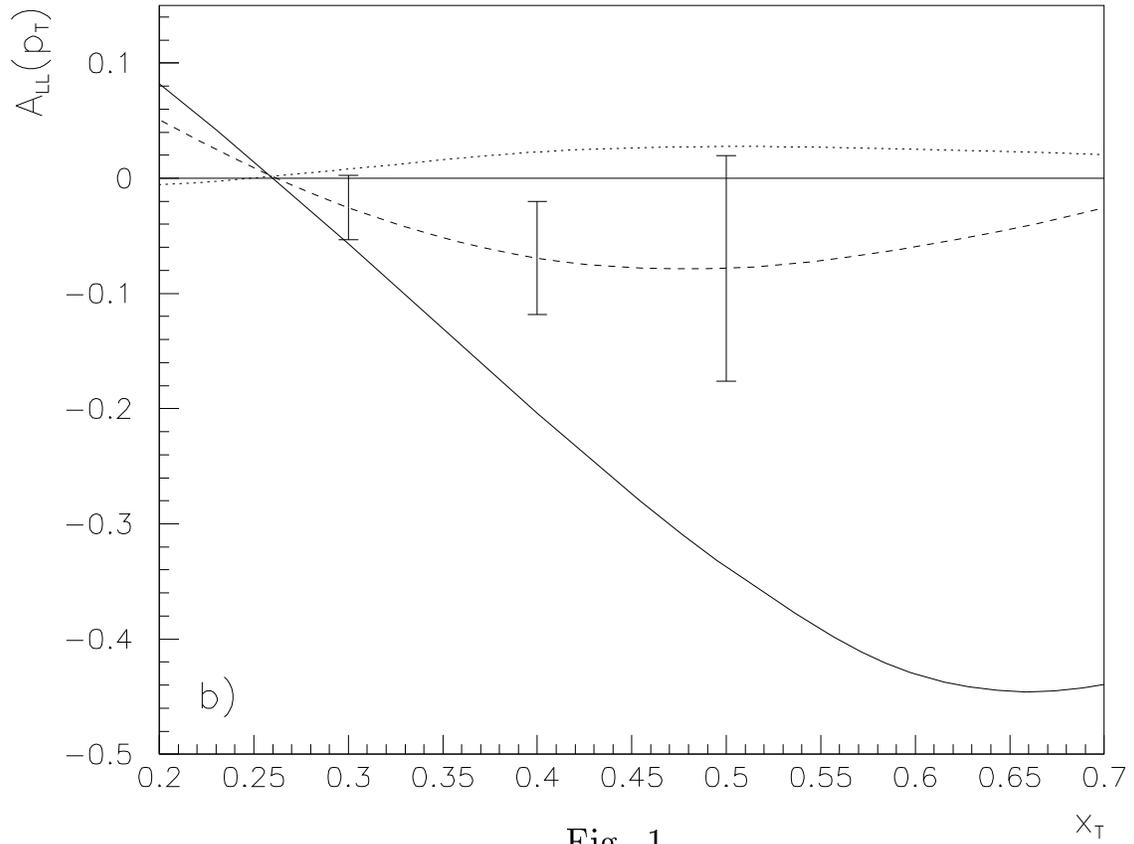
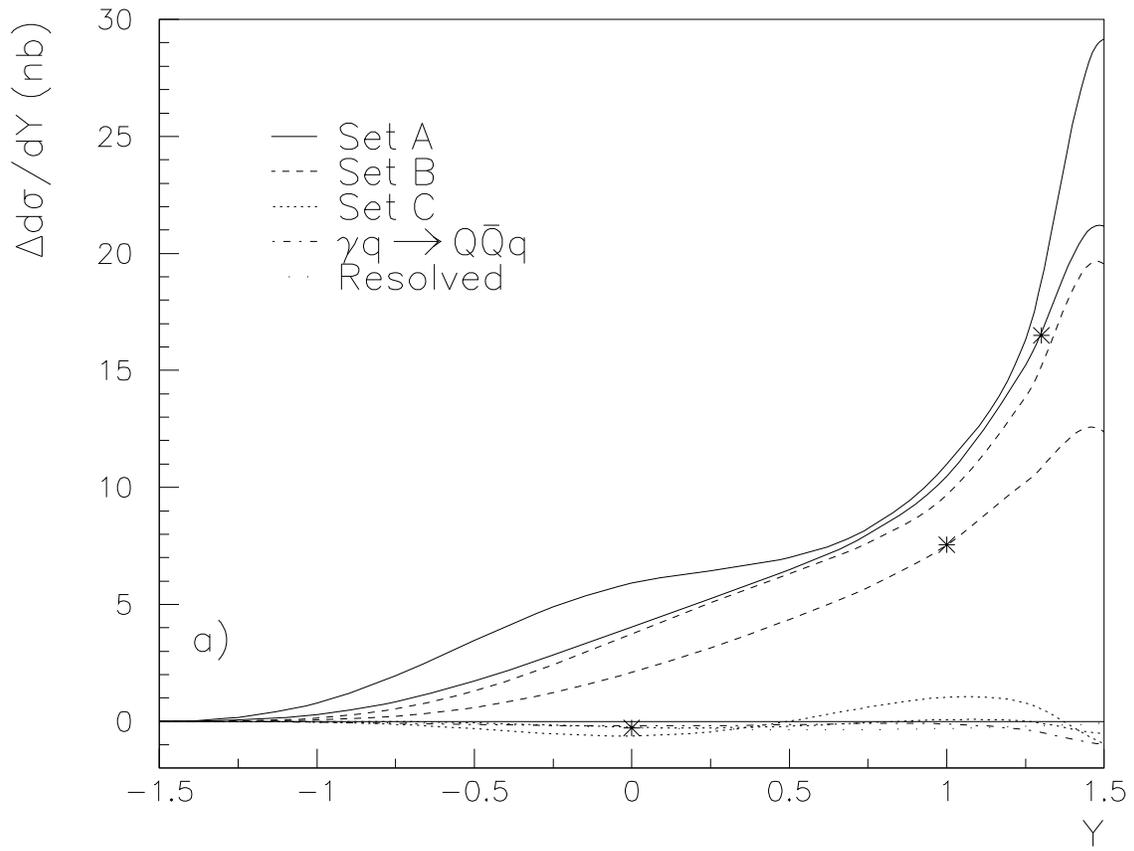
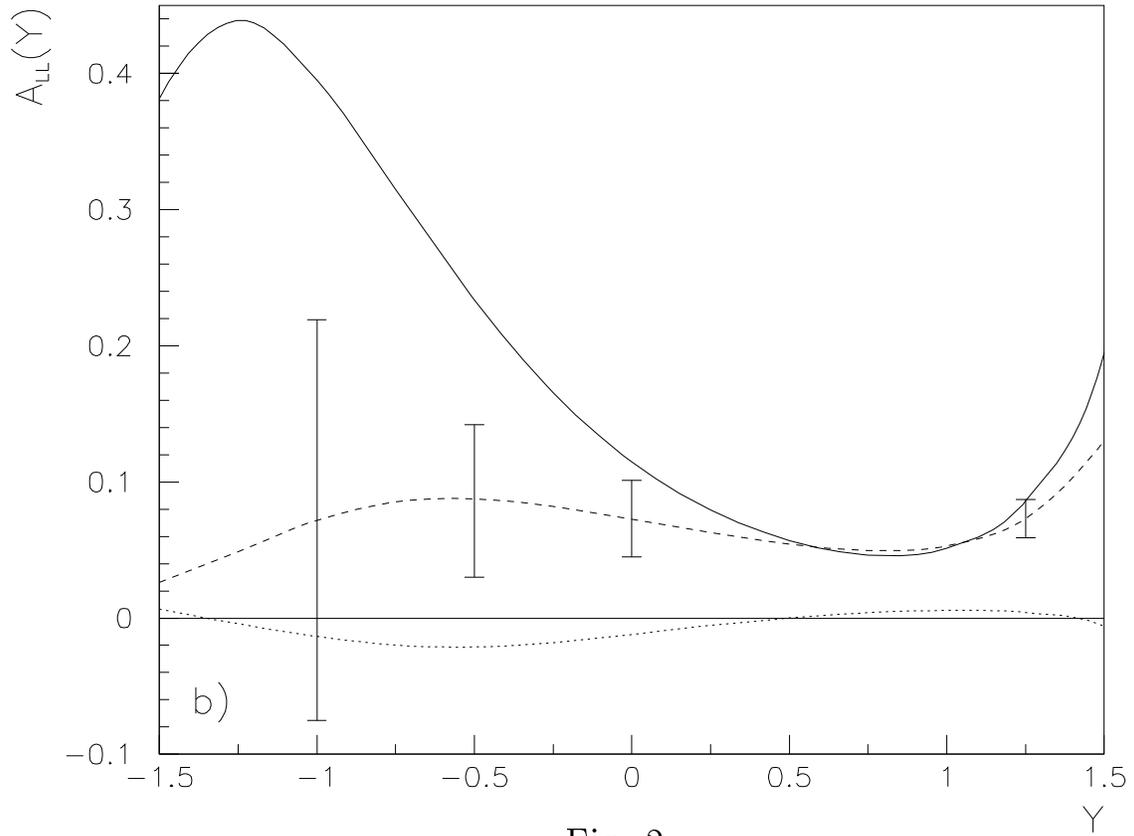
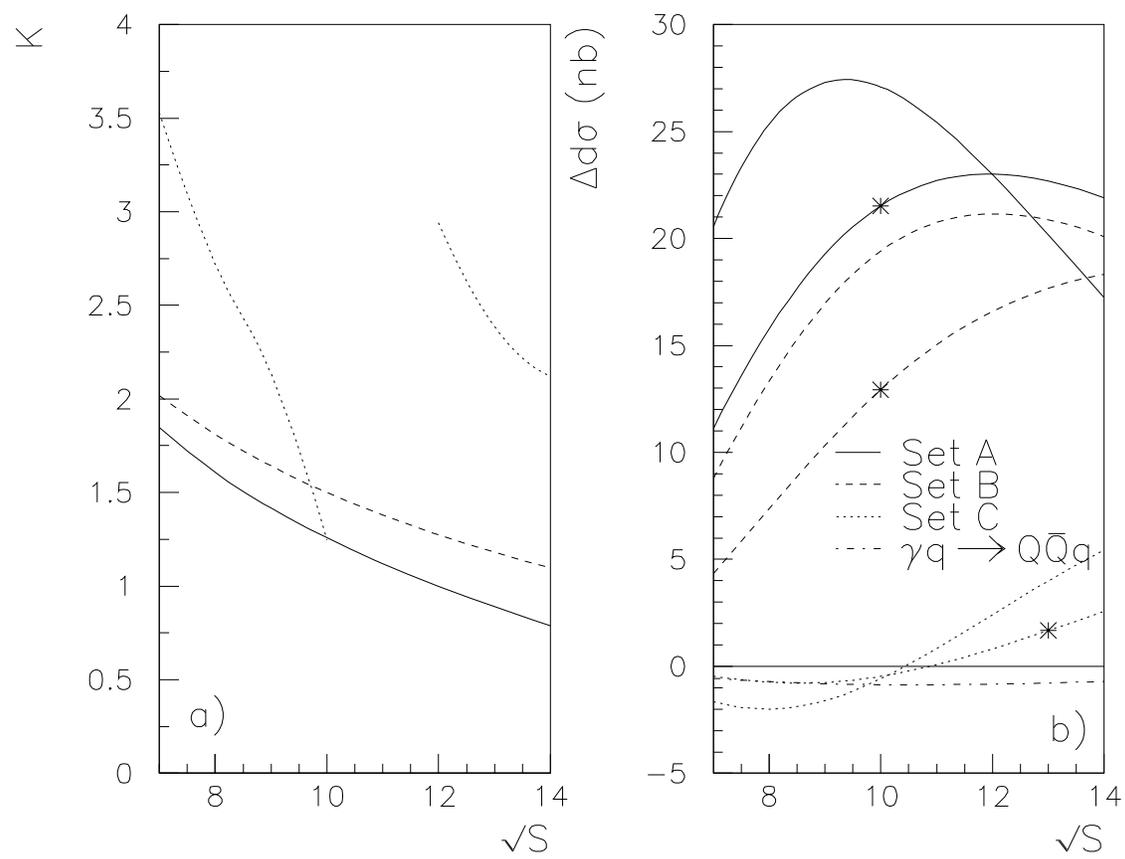
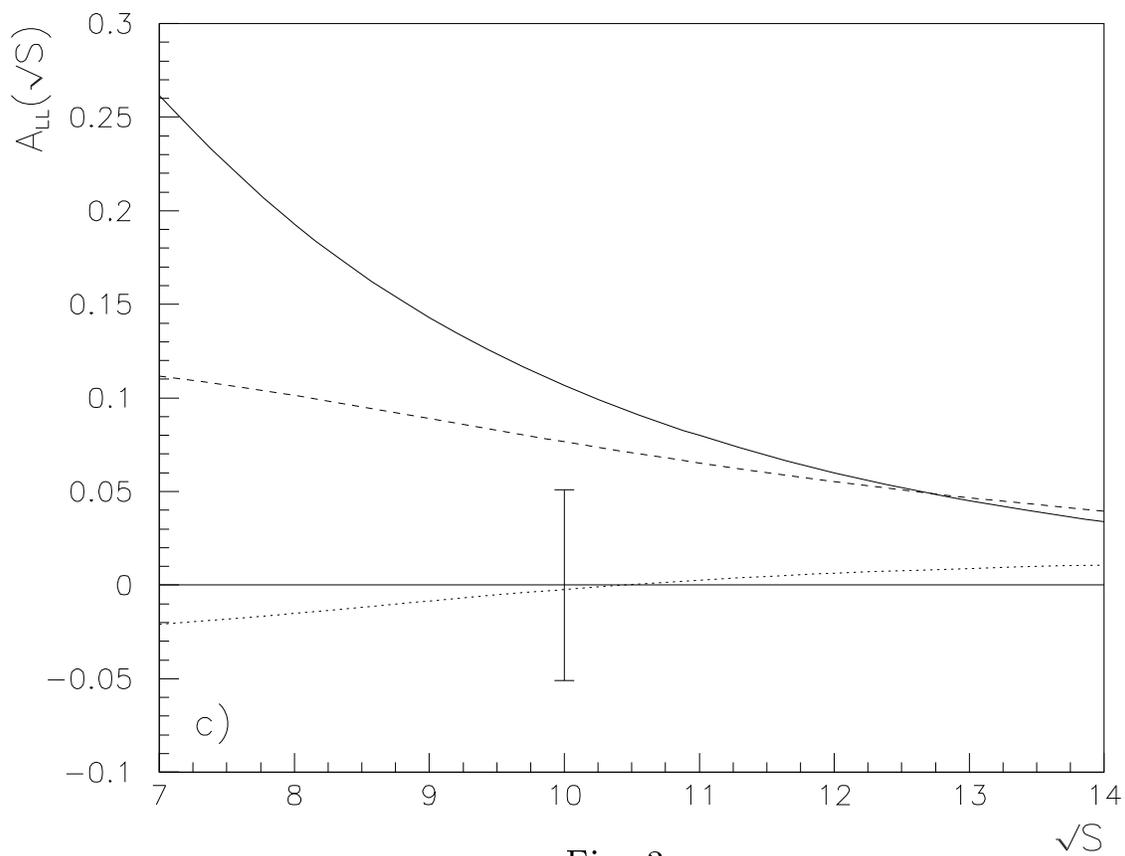
\end{document}